\documentclass{article}
\usepackage{xcolor}

\usepackage[english]{babel}
\usepackage{amssymb}
\usepackage{authblk}

\usepackage[letterpaper,top=2cm,bottom=2cm,left=3cm,right=3cm,marginparwidth=1.75cm]{geometry}

\usepackage{amsmath}
\usepackage{graphicx}
\usepackage[colorlinks=true, allcolors=blue]{hyperref}

\title{Bounds of Block Rewards in Honest PinFi Systems}

\author[1]{Qi~He}
\author[1]{Yunwei~Mao}
\author[2,*]{Ju~Li}
\affil[1]{Loopro Inc., Cambridge, MA 02139, USA (\href{https://www.loopro.ai/}{https://www.loopro.ai/})}
\affil[2]{Massachusetts Institute of Technology, Cambridge, MA 02139, USA}
\affil[*]{has never owned and will never own any LooPIN tokens.}
\date{\today}

\begin{document}
\maketitle

\begin{abstract}
PinFi is a class of novel protocols for decentralized pricing of dissipative assets, whose value naturally declines over time. Central to the protocol's functionality and its market efficiency is the role of liquidity providers (LPs). This study addresses critical stability and sustainability challenges within the protocol, namely: the propensity of LPs to prefer selling in external markets over participation in the protocol; a similar inclination towards selling within the PinFi system rather than contributing as LPs; and a scenario where LPs are disinclined to sell within the protocol. Employing a game-theoretic approach, we explore PinFi's mechanisms and its broader ramifications. Our findings reveal that, under a variety of common conditions and with an assumption of participant integrity, PinFi is capable of fostering a dynamic equilibrium among LPs, sellers, and buyers. This balance is maintained through a carefully calibrated range of block rewards for LPs, ensuring the protocol's long-term stability and utility.
\end{abstract}

\section{Introduction}

Classical Automatic Market Maker (AMM) protocols, such as Logarithmic Market Scoring Rules (LMSRs) \cite{hanson2003combinatorial}, the Bancor protocol \cite{hertzog17benartzi, hertzog18benartzi}, and Uniswap's constant product protocol \cite{zhang18chen, adams2020uniswap}, have been instrumental in shaping asset pricing within the Decentralized Finance (DeFi) ecosystem. Each of these protocols offers a unique approach to asset valuation: LMSRs are optimized for forecasting discrete event outcomes; the Bancor protocol uses a bonding curve to match an asset's price to its supply-driven equilibrium price; and Uniswap relies on reserve balances for pricing, independent of supply adjustments.

However, these protocols primarily cater to non-dissipative assets, whose value remains unaffected by time. This limitation is evident when considering the complexity of transactions involving dissipative-to-non-dissipative asset pairs, such as those in markets for GPU cloud computing, electricity and other perishable resource (e.g. fruits and vegetables) distribution, or services — areas where traditional AMMs fall short. Addressing this gap, the PinFi protocol \cite{mao2024} introduces a protocol for the decentralized pricing of dissipative assets. Drawing inspiration from Uniswap's widely adopted constant product mechanism, PinFi adapts this principle to suit the fluctuating nature of dissipative assets, thereby extending the applicability of AMM principles to a new asset class.

Despite the acknowledged potential of PinFi, detailed explorations of its operational intricacies and broader market impact are scarce. At the heart of PinFi's design is the role of liquidity providers (LPs), who are incentivized to stabilize the market through resource staking, thereby facilitating balanced transactions for buyers and sellers. The economic viability of PinFi hinges on maintaining a delicate balance among LPs, buyers, sellers, and external market dynamics. Three primary challenges threaten this balance: insufficient incentives leading LPs to exit for external markets; the allure of arbitrage opportunities within PinFi overshadowing long-term participation rewards; and overly generous block rewards skewing participation heavily towards liquidity provision. These potential vulnerabilities necessitate a rigorous analysis to ensure the protocol's stability and longevity.

This paper aims to bridge this gap by a game-theoretic analysis of PinFi, evaluating how it achieves equilibrium among LPs, buyers, and sellers under a spectrum of conditions and assuming honest participation. Such equilibrium is crucial for ensuring fair pricing of dissipative assets and the overall health of the protocol.

The remainder of this paper is structured as follows: Section \ref{honestsystem} outlines the foundational principles of the PinFi system, predicated on the integrity of its participants. Section \ref{simplecases} examines the dynamics within these systems, focusing on limited participant scenarios to understand the conditions for equilibrium. Finally, Section \ref{ongoingwork} summarizes our findings and suggests avenues for future research.

\section{Honest PinFi Systems With Resource Speculators}
\label{honestsystem}

\begin{figure}[ht]
\centering
\includegraphics[width=1.0\linewidth]{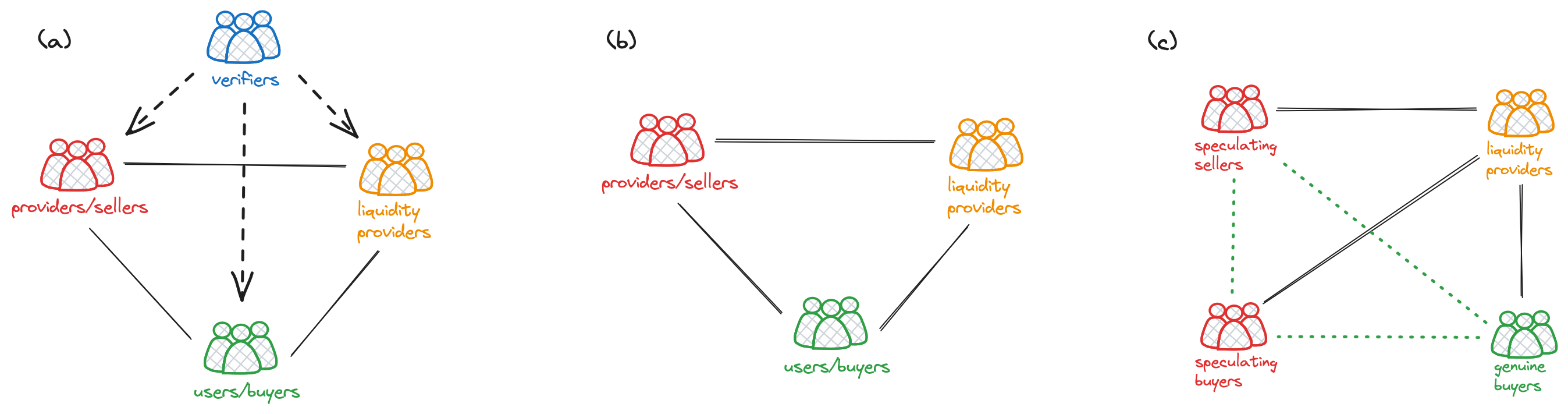}
\caption{\label{fig:scheme}The dynamics of the PinFi system are explored through (a) a general framework and (b) a framework assuming honesty among participants. In (c), participants are categorized by their economic motivations: speculating sellers, speculating buyers, genuine buyers, and liquidity providers.}
\end{figure}

The PinFi framework, introduced by our previous manuscript \cite{mao2024}, identifies four key roles within its ecosystem based on their relationship to resources and/or duties, as illustrated in Figure \ref{fig:scheme}(a):

\begin{itemize}
\item {\bf{Liquidity Providers}}: These are participants who contribute both dissipative and an equal value of non-dissipative assets to enhance the pool's liquidity. Upon the expiration of the dissipative asset, they receive rewards for their contribution and are returned the non-dissipative assets at the current exchange rate.

\item {\bf{Holders/Sellers}}: Holders in this category choose to sell their dissipative assets to the pool in exchange for non-dissipative assets. They are subject to two types of fees: a staking fee for generating an on-chain dissipative certificate, and an exchange fee incurred when withdrawing liquidity, which is then burned.

\item {\bf{Users/Buyers}}: These are the end-users, such as clients, developers, or companies, in need of the dissipative assets for their intrinsic value, such as AI companies in need of GPU computing hours. While the PinFi system facilitates these transactions, users also have the option to acquire these assets externally. For example, users can always purchase GPU computing hours from services like AWS, Google Cloud, or Azure.

\item {\bf{Verifiers}}: To ensure the integrity of transactions and mitigate the risk of dishonesty among participants, verifiers play a critical role. They employ a decentralized, secure protocol (such as proofs-of-computing-power-staking in the original PinFi paper) for verification, addressing Byzantine faults and ensuring the system's reliability.
\end{itemize}

In the PinFi ecosystem, both {\bf{Liquidity Providers}} and {\bf{Providers/Sellers}} are miners who possess dissipative assets. The key distinction lies in their approach to these assets: Providers/Sellers prefer to quickly exchange their dissipative assets for non-dissipative ones without engaging in staking. On the other hand, Liquidity Providers are willing to stake both their dissipative and non-dissipative assets in liquidity pools, despite the potential for impermanent loss, a common characteristic of Automatic Market Makers (AMMs). As compensation for their contribution and to mitigate the risks of impermanent loss, Liquidity Providers are rewarded with block rewards in newly issued blocks.

In this analysis, we operate under the assumption of universal honesty among PinFi system participants. This encompasses Providers (encompassing both Sellers and Liquidity Providers), who are assumed to reliably supply the required dissipative assets to users, and Users, who are presumed to accurately confirm the efficacy of acquired dissipative assets. Given this premise of inherent honesty, the role of Verifiers becomes redundant, as the verification process invariably yields a truthful affirmation. Consequently, a simplified representation of the system, as depicted in Figure \ref{fig:scheme}(b), is allowed.

Economic systems inherently include speculators. For instance, when the price of a dissipative asset in the PinFi liquidity pool significantly exceeds external market rates, resource-holding speculators emerge as sellers, driving the asset's price down within the PinFi system. Conversely, if the PinFi pool's asset price is substantially lower, speculators act as buyers, purchasing assets to sell externally for a profit, thus pushing the PinFi asset price up. While both speculating sellers and buyers aim to exploit arbitrage opportunities for profit, genuine buyers seek the resources themselves, with price being their primary concern. Genuine buyers will participate in the PinFi system as long as its prices are competitive with external markets. The interactions among speculating sellers, speculating buyers, genuine buyers, and liquidity providers significantly influence the PinFi system's dynamics. We categorize system participants based on their economic interests:
\begin{itemize}
\item {\bf{Speculating Sellers (SSs)}}: Individuals holding resources who become sellers when the PinFi liquidity pool's price for a dissipative asset significantly surpasses that of external markets.

\item {\bf{Speculating Buyer (SBs)}}: Speculators who buy assets from the PinFi pool at lower prices to sell them externally for a profit.

\item {\bf{Genuine Buyers (GBs)}}: Users who engage with the PinFi system, motivated by the competitive pricing of its assets compared to external markets.

\item {\bf{Liquidity Providers (LPs)}}: Participants who contribute their resources to the liquidity pool in exchange for block rewards.
\end{itemize}
These classifications and their interactions within the PinFi system are illustrated in Figure \ref{fig:scheme}(c).

\section{Dynamics Between Participants: Quasi-static Cases}
\label{simplecases}

Despite the presumption of honesty within the PinFi system, the interplay among the different participant types introduces complexity. This section delves into the dynamics and potential equilibria in simplified scenarios, focusing on interactions between just two participant types in the honest PinFi framework.

Building upon the foundational understanding of the economic interactions within the network's liquidity pool, let us contextualize these quantities and assumptions to streamline the subsequent analysis. This approach will help clarify the framework within which speculative trading and liquidity provision operate, setting the stage for a deeper mathematical exploration. In analyzing the dynamics of the network’s liquidity pool, we consider the following essential parameters:

\begin{itemize}
\item {\bf{Initial Liquidity Pool State}}: The pool starts with a stock of dissipative assets, denoted as $\mathcal{N}$ (measured in power*hours), and an equivalent value of non-dissipative assets, valued at $\alpha\mathcal{N}$ USD. The liquidity providers will not remove liquidity from the pool or add liquidity to the pool during the process. 
\item {\bf{Total Seller Fees}}: Sellers incur a combined cost of $\beta$ USD per power*hour, encompassing both the staking and exchange fees. This fee influences the sellers’ willingness to engage in trading within the pool.
\item {\bf{LP Rewards}}: The rewards for liquidity providers are represented by $\gamma$ USD per power*hour, incentivizing the provision of liquidity to the pool.
\item {\bf{Buyer Exchange Fee}}: Buyers face an exchange fee denoted by $\delta$ USD per power*hour, affecting their purchasing decisions within the ecosystem.
\item {\bf{External Service Cost}}: This represents the expense of acquiring similar services from external providers, denoted by $\theta$ USD per power hour. It serves as a benchmark for assessing the competitiveness of the network's pricing. When the price is set at $\theta$, the fill rate is defined as $p<1$. Although a more detailed fill rate model could be applied in numerical simulations, a simplified version is utilized here for theoretical analysis.
\end{itemize}
To make our analysis possible, further assumptions are made:
\begin{itemize}
\item {\bf{Staking Fee for LPs}}: The analysis simplifies by omitting direct staking fees for liquidity providers, focusing instead on the rewards and potential gains from their participation.
\item {\bf{Operational Costs}}: The costs $c$ associated with providing the service, including electronics and management, are encapsulated in the $\alpha$ pricing. This assumption simplifies the economic model by integrating operational costs into the initial asset valuation.
\item {\bf{Stability in Non-dissipative Asset Pricing}}: We assume that the price of non-dissipative assets, when measured against the USD, remains constant throughout the analysis period. This assumption eliminates the complexity that would arise from fluctuating prices, allowing us to focus on the effects of other variables and interactions within the liquidity pool without the added variable of price volatility for these assets.
\item {\bf{Quasi-static Condition}}: The depreciation rate of the dissipative asset significantly outpaces the response rate of any system participant, allowing us to disregard interactions occurring at differing frequencies.

\end{itemize}

With these parameters and assumptions in place, we are equipped to delve into the economic mechanisms underpinning the PinFi network, assessing the interactions between speculating sellers and liquidity providers (SS-LP), speculating buyers and liquidity providers (SB-LP), and genuine buyers and liquidity providers (GB-LP). 

\subsection{Dynamics Between Speculating Sellers and Liquidity Providers}
The exchange dynamics within liquidity pools involve both dissipative assets (like computing power, which depletes over time) and non-dissipative assets. Dissipative assets are priced at $\alpha$ USD per power hour, with a selling cost of $\beta$. This pricing structure incentivizes speculating sellers to exchange their dissipative assets for non-dissipative ones whenever $p\theta < \alpha-\beta$. For each unit of dissipative asset sold, a speculating seller earns $\alpha-\beta-p\theta$ more than they would in the external market. The greater the value of $\alpha-\beta-p\theta$, the more motivated a speculating seller becomes to trade her resources in the PinFi liquidity pool. SSs exit the system upon the fulfillment of the condition $\alpha = \beta + p\theta$.

Consider a scenario at time $t$ where the quantity of the dissipative asset contributed by liquidity providers to the system is represented by $\hat{\mathcal{N}}(t)$, and the exchange rate from non-dissipative to dissipative assets is $\hat{\alpha}(t)$. Thus, the liquidity pool's state is given by the vector $(\hat{\mathcal{N}}(t), \hat{\alpha}(t)\hat{\mathcal{N}}(t))$. At the outset, a speculating seller generates a dissipative asset, representing a service lasting for $\Delta t$, and sells this asset to the liquidity pool. The post-trade state of the liquidity pool, governed by the constant product formula, becomes:

\begin{equation}
\left(\hat{\mathcal{N}}(t)+\Delta t, \frac{\hat{\mathcal{N}}(t)}{\hat{\mathcal{N}}(t)+\Delta t}\hat{\alpha}(t)\hat{\mathcal{N}}(t)\right), 
\end{equation}
yielding a profit for the speculating seller of $\dot{\mathcal{P}}\Delta t = (\hat{\alpha}(t) - \beta)\Delta t$, which simplifies to $\dot{\mathcal{P}} = \hat{\alpha}(t) - \beta$.

Under the quasi-static condition and in the absence of buyers, the dissipative asset $\Delta$ is considered expired immediately post-trade. This results in the liquidity pool reverting to:
\begin{equation}
\left(\hat{\mathcal{N}}(t), \frac{\hat{\mathcal{N}}(t)}{\hat{\mathcal{N}}(t)+\Delta t}\hat{\alpha}(t)\hat{\mathcal{N}}(t)\right).
\end{equation}
Consequently, the post-trade exchange rate for non-dissipative assets, after the traded asset's expiration, is defined as:
\begin{equation}
\hat{\alpha}(t+\Delta t)  = \frac{\hat{\mathcal{N}}(t)}{\hat{\mathcal{N}}(t)+\Delta t}\hat{\alpha}(t).
\label{eq1}
\end{equation}
Assuming liquidity providers' contributions remain unchanged, the dynamics within the liquidity pool are attributed solely to liquidity withdrawals by sellers and the expiration of unused dissipative assets, simplifying  $\hat{\mathcal{N}}(t)$ to a constant $\mathcal{N}$. This assumption leads to the differential equation for the exchange rate and its solution:
\begin{equation}
\frac{\mathrm{d}\ln\hat{\alpha}(t)}{\mathrm{d}t} = -\frac{1}{\mathcal{N}}, \ \ \ \hat{\alpha}(t) = \alpha \exp(-t/\mathcal{N}).
\end{equation}
The cessation time (T), or the point at which speculation becomes unprofitable ($\hat{\alpha}(t) \equiv \beta +p\theta$), is determined by:
\begin{equation}
T = \mathcal{N}\ln(\alpha/(\beta+p\theta)).
\label{texpress1}
\end{equation}

We now assess the net gains for both the speculating seller and the liquidity provider within the specified timeframe:
\begin{enumerate}
\item For the speculating seller. The net gain, denoted as $\mathcal{G}_{\text{\tiny ss}}$, is derived as follows:
\begin{equation}
\mathcal{G}_{\text{\tiny ss}} = \int_0^T\dot{\mathcal{P}}\mathrm{d}t =\mathcal{N}(\beta+p\theta) \left[\frac{\alpha}{\beta+p\theta} - \frac{\beta}{\beta+p\theta} \ln\left(\frac{\alpha}{\beta+p\theta}\right)-1\right].
\end{equation}

\item For the liquidity provider. The liquidity provider's profit, $\mathcal{G}_{\text{\tiny lp}}$, combines earned block rewards, the impact of impermanent loss at cessation time, and the opportunity cost/loss. The accumulated rewards, $\mathcal{A} = \gamma\mathcal{N} T$, and the realized impermanent loss $\mathcal{L} = (\beta+p\theta - \alpha)\mathcal{N}$, yield:
\begin{equation}
\mathcal{G}_{\text{\tiny lp}} = \mathcal{A} +\mathcal{L} = \mathcal{N}(\beta +p\theta)\left[1+\frac{\gamma \mathcal{N}}{\beta+p\theta}\ln\left(\frac{\alpha}{\beta+p\theta}\right) - \frac{\alpha}{\beta+p\theta}
\right].
\end{equation}
\end{enumerate}
The existence of the liquidity pool hinges on the presence of liquidity providers. The opportunity for speculating sellers to exchange their dissipative assets for non-dissipative ones, even in the absence of buyers, is made possible solely through the liquidity provided by these providers. When the disparity between $\alpha - \beta$ and $p\theta$ is significant and block rewards $\gamma$ are minimal, achieving equilibrium between speculating sellers and liquidity providers becomes untenable. Rational liquidity providers would abandon their roles in favor of becoming speculating sellers to capitalize on the immediate, higher gains, overshadowing the lesser block rewards.

Conversely, even with a large disparity between $\alpha - \beta$ and $p\theta$, if block rewards $\gamma$ substantially exceed the arbitrage gains, rational resource holders will not opt to become speculating sellers. Instead, they are incentivized to remain as liquidity providers, attracted by the higher rewards, despite potential long-term depreciation in the non-dissipative asset's value due to inflation.

Therefore, intuitively, there must exist a balanced relationship between $\alpha$ and $\gamma$, given the parameters $(\beta, p, \theta)$, that motivates some participants to act as speculating sellers while others prefer the role of liquidity providers. Ideally, the economic equilibrium is reached when the earnings from speculating selling during arbitrage are equivalent to the rewards accrued by liquidity providers in the same period, i.e., $\mathcal{G}_{\text{\tiny ss}} = \frac{1}{\mathcal{N}}\mathcal{G}_{\text{\tiny lp}}$. This leads to the following equation:
\begin{equation}
\frac{\gamma}{\beta+p\theta} = \left(1 + \frac{1}{\mathcal{N}}\right)\frac{\left(\frac{\alpha}{\beta+p\theta}-1\right)}{\ln\left(\frac{\alpha}{\beta+p\theta}\right)}-\left(\frac{\beta}{\beta +p\theta}\right) , \, \, \text{for }\, \frac{\alpha}{\beta+p\theta}\geq 1.
\label{lp_seller_eq_normalized}
\end{equation}
The normalized profit for either speculating sellers or liquidity providers, $\hat{\mathcal{G}}_{\text{\tiny sl}}$, is given by:
\begin{equation}
\hat{\mathcal{G}}_{\text{\tiny sl}} = \frac{\mathcal{G}_{\text{\tiny ss}}}{\mathcal{N}(\beta + p\theta)} = \frac{\alpha}{\beta+p\theta} - \frac{\beta}{\beta+p\theta} \ln\left(\frac{\alpha}{\beta+p\theta}\right)-1.
\label{gsl_normalized}
\end{equation}

\begin{figure}
\centering
\includegraphics[width=0.5\linewidth]{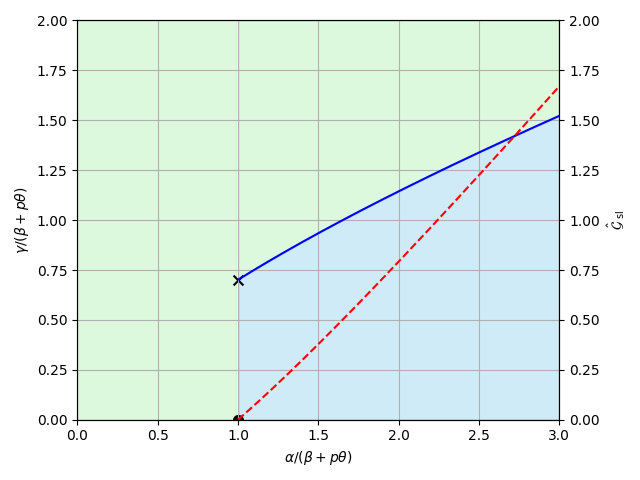}
\caption{\label{fig:lp_seller}The dynamics between the SSs and LPs.}
\end{figure}

Figure \ref{fig:lp_seller} visualizes the equilibrium between SSs and LPs, delineated by a blue line, with parameter $\frac{\beta}{\beta+p\theta} = 0.3$ and $\mathcal{N} = 1000$. This line divides the parameter space into two distinct zones: the sky blue zone $\mathcal{SS}$, indicating a preference for SSs, and the light green zone $\mathcal{LP}$, signifying a predilection towards being an LP. The equilibrium line exhibits a monotonic increase across the phase diagram, implying that an increase in $\frac{\alpha}{\beta + p\theta}$ necessitates a corresponding rise in the normalized block rewards $\frac{\gamma}{\beta + p\theta}$.  Another notable feature within the diagram is a vertical line, spanning from a black cross to a black dot, signifying the equilibrium point at the initial setup where no arbitrage opportunities exist.

Accompanying the equilibrium delineation, the diagram also includes an earnings curve, represented by a red dashed line. It becomes apparent that when the initial setup satisfies $\frac{\alpha}{\beta + p\theta} \leq 1$, the normalized profit is zero. Conversely, as $\frac{\alpha}{\beta + p\theta} > 1$, the normalized profit exhibits a monotonic increase in conjunction with the initial asset pricing, $\alpha$.

As we conclude this section, we delve into the limit case characterized by $\frac{\alpha}{\beta+p\theta}$ approaching just above 1. In this scenario, the adjusted block rewards for speculating sellers, denoted as $\left.\frac{\gamma}{\beta+p\theta}\right|_{\frac{\alpha}{\beta+p\theta}\rightarrow 1^+}^{\text{\tiny ss}}$, are formulated as:
\begin{equation}
\left.\frac{\gamma}{\beta+p\theta}\right|_{\frac{\alpha}{\beta+p\theta}\rightarrow 1^+}^{\text{\tiny ss}} = 1+\frac{1}{\mathcal{N}} - \frac{\beta}{\beta+p\theta}.
\label{eq:seller_limit_gamma}
\end{equation}
This expression encapsulates the dynamics of block rewards in proximity to this critical threshold. Moreover, the normalized profit for speculating sellers, as $\frac{\alpha}{\beta+p\theta}$ marginally exceeds 1, is determined to be:
\begin{equation}
\left.\hat{\mathcal{G}}_{\text{\tiny sl}}\right|_{\frac{\alpha}{\beta+p\theta}\rightarrow 1^+} = 0.
\label{eq:seller_limit_G}
\end{equation}

\subsection{Dynamics Between Buyers and Liquidity Providers}

The system categorizes buyers into two distinct types: speculating buyers (SBs) and genuine buyers (GBs). Speculating buyers are engaged in purchasing assets from liquidity pools and selling them in external markets, driven by the potential for profit. Their participation is contingent upon the viability of arbitrage opportunities; the absence of guaranteed profits prompts their exit from the system. The cost incurred by SBs when buying from the liquidity pool is denoted as $\alpha + \delta$, while the return from selling these acquired dissipative assets to external sources is represented by $p\theta$. Thus, the condition signaling the withdrawal of speculating sellers is expressed as:
\begin{equation}
\alpha +\delta \leq p\theta.
\end{equation}
Conversely, genuine buyers are primarily focused on acquiring resources for their projects efficiently, opting to purchase dissipative assets through the Automated Market Maker (AMM) system. Their participation threshold is determined by the condition:
\begin{equation}
\alpha + \delta \leq \theta.
\label{eq:gbcondition}
\end{equation}
This equation delineates the maximum price at which GBs are willing to engage in the system. Consequently, as arbitrage opportunities diminish, SBs are the first to exit the system when $\alpha + \delta = p\theta$. Following this, GBs continue to exert upward pressure on the price until it reaches the point where $\alpha + \delta = \theta$, at which GBs cease their purchases of the dissipative assets.

Consider the moment at time $t$, with the liquidity pool defined by the vector $(\hat{\mathcal{N}}(t), \hat{\alpha}(t)\hat{\mathcal{N}}(t))$. A buyer's acquisition of $\Delta t$ power hours alters the pool's state as dictated by the constant product formula to:
\begin{equation}
\left(\hat{\mathcal{N}}(t)-\Delta t, \hat{\alpha}(t)\frac{\hat{\mathcal{N}}(t)}{\hat{\mathcal{N}}(t)-\Delta t}\hat{\mathcal{N}}(t)\right)
\end{equation}
Under the quasi-static condition, which assumes immediate replenishment of dissipative assets by liquidity providers at $t + \Delta t$, the pool's configuration transitions to:
\begin{equation}
\left(\hat{\mathcal{N}}(t), \hat{\alpha}(t)\frac{\hat{\mathcal{N}}(t)}{\hat{\mathcal{N}}(t)-\Delta t}\hat{\mathcal{N}}(t)\right)
\end{equation}
This adjustment yields a revised exchange rate, $\hat{\alpha}(t+\Delta t) = \hat{\alpha}(t)\frac{\hat{\mathcal{N}}(t)}{\hat{\mathcal{N}}(t)-\Delta t}$, , and introduces the following dynamic equation for the exchange rate $\frac{\mathrm{d}\ln \hat{\alpha}(t)}{\mathrm{d}t} = \frac{1}{\hat{\mathcal{N}}(t)}$.
With the assumption that contributions from liquidity providers remain consistent over time, changes within the liquidity pool are driven exclusively by the withdrawal of liquidity by sellers and the natural expiration of unused dissipative assets. This allows us to treat $\hat{\mathcal{N}}(t)$ as a constant, denoted by $\mathcal{N}$. Consequently, the differential equation governing the exchange rate $\hat{\alpha}(t)$ simplifies to:
\begin{equation}
\frac{\mathrm{d}\ln \hat{\alpha}(t)}{\mathrm{d}t} = \frac{1}{\mathcal{N}}, \,\,\,\hat{\alpha}(t) = \alpha\exp(t/\mathcal{N}).
\end{equation}
The cessation time (T), or the point at which speculation becomes unprofitable for the speculating buyers ($\hat{\alpha}(T) \equiv p\theta -\delta$), is determined by:
\begin{equation}
T = \mathcal{N}\ln((p\theta-\delta)/\alpha).
\label{texpress2}
\end{equation}

The sustainability of the liquidity pool is critically dependent on the engagement of liquidity providers (LPs). The feasibility for speculating buyers (SBs) to purchase dissipative assets at lower prices from the PinFi system and subsequently sell them at a higher price externally is predicated on the liquidity furnished by LPs. In scenarios where block rewards, $\gamma$, are insufficient, rational LPs may opt to exit their roles, preferring to directly sell their resources in external markets. The potential gain for these rational LPs, denoted as $\mathcal{G}_{\text{\tiny lp}}^{\text{\tiny rational}}$, is calculated by integrating the profit, $p\theta\mathcal{N}$, over the cessation period, resulting in:
\begin{equation}
\begin{split}
\mathcal{G}_{\text{\tiny lp}}^{\text{\tiny rational}} &= \int_0^T p\theta\mathcal{N}\mathrm{d}t =p\theta\mathcal{N}^2\ln\left(\frac{p\theta-\delta}{\alpha}\right).
\end{split}
\end{equation}
For actual LPs, the profit, $\mathcal{G}_{\text{\tiny lp}}$, encompasses both the earned block rewards, $\mathcal{A} = \gamma \mathcal{N}T$, and the impact of impermanent loss at the cessation, $\mathcal{L} = \alpha(T)\mathcal{N}(T) - \alpha \mathcal{N}$, expressed as:
\begin{equation}
\begin{split}
\mathcal{G}_{\text{\tiny lp}} &= \mathcal{A} + \mathcal{L}=\gamma\mathcal{N}^2\ln\left(\frac{p\theta-\delta}{\alpha}\right) + (p\theta-\delta-\alpha)\mathcal{N}.
\end{split}
\end{equation}

A balanced relationship between $\alpha$ and $\gamma$, influenced by the parameters $(\beta, p, \theta)$, is imperative to motivate LPs to remain within the liquidity pool. The economic equilibrium is ideally achieved when the opportunity loss of LPs for staying in the pool, $\mathcal{G}{\text{\tiny lp}}^{\text{\tiny rational}}$, equates to the gains obtained by LPs for their role, $\mathcal{G}{\text{\tiny lp}}$, leading to:
\begin{equation}
\frac{\gamma}{\beta+p\theta} = -\frac{1}{\mathcal{N}}\frac{\left(\frac{p\theta-\delta}{\beta+p\theta}-\frac{\alpha}{\beta+p\theta}\right)}{\ln\left(\frac{p\theta-\delta}{\beta+p\theta}\right)-\ln\left(\frac{\alpha}{\beta+p\theta}\right)}+\left(1-\frac{\beta}{\beta+p\theta}\right), \, \, \text{for }\, \frac{\alpha}{\beta+p\theta}\leq \frac{p\theta-\delta}{\beta+p\theta}.
\label{lp_buyer_eq_normalized}
\end{equation} 
The normalized profit (either rational LPs or actual LPs) $\mathcal{G}_{\text{\tiny bl}}$is
\begin{equation}
\hat{\mathcal{G}}_{\text{bl}} = \frac{\mathcal{G}_{\text{\tiny lp}}^{\text{\tiny rational}}}{\mathcal{N}^2(\beta + p\theta)} = \left(1-\frac{\beta}{\beta+p\theta}\right)\left[\ln\left(\frac{p\theta-\delta}{\beta+p\theta}\right) -\ln\left(\frac{\alpha}{\beta+p\theta}\right) \right].
\label{gbl_normalized}
\end{equation}

\begin{figure}[ht]
\centering
\includegraphics[width=1.0\linewidth]{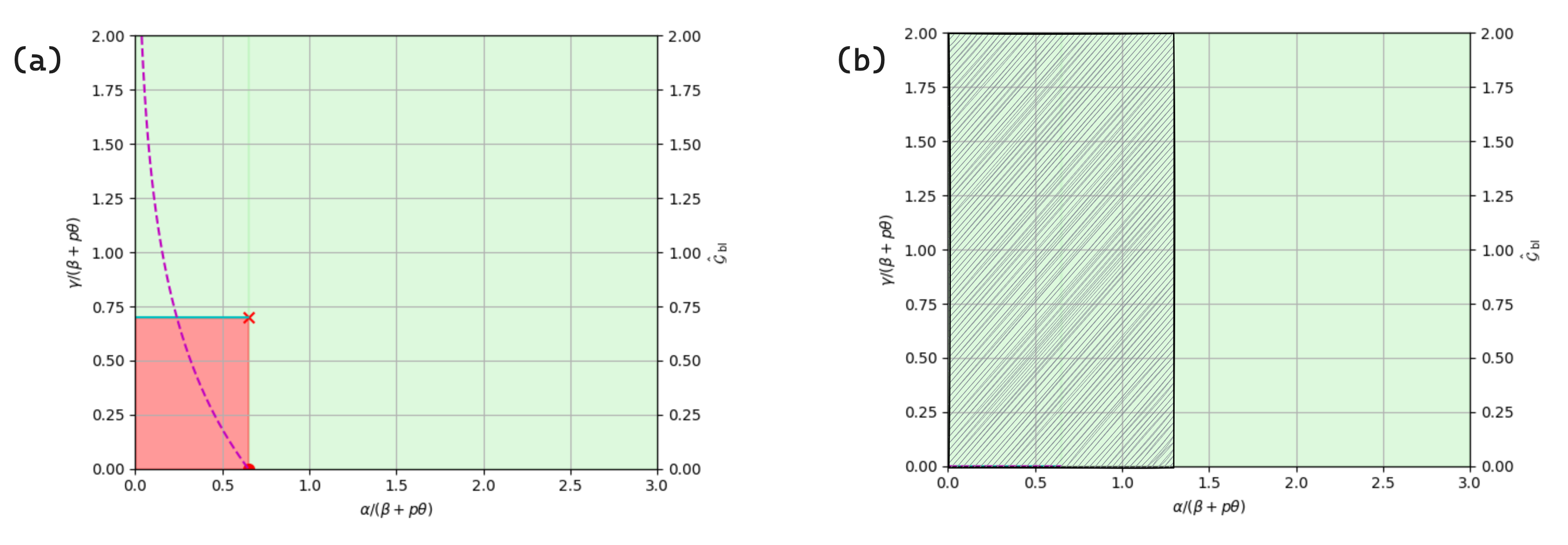}
\caption{\label{fig:lp_buyer}The dynamics between liquidity providers and the (a) speculating buyers and (b) genuine buyers.}
\end{figure}

Figure \ref{fig:lp_buyer}(a) presents the equilibrium and profit dynamics between rational liquidity providers (LPs)—indicative of the presence of speculating buyers (SBs)—and LPs within the PinFi system, utilizing parameter ratios $\frac{\beta}{\beta+p\theta} = 0.3$, $\frac{p\theta-\delta}{\beta+p\theta} = 0.65$, and $\frac{\theta-\delta}{\beta+p\theta}=1.3$. The equilibrium state is marked by a cyan line, which bifurcates the parameter space into two distinctive regions: a red zone ($\mathcal{SB}$), highlighting the propensity of LPs to exit the system and directly sell resources in external markets; and a light green zone ($\mathcal{LP}$), indicating conditions favorable for participation as an LP within the pool. A significant diagrammatic feature is a red cross that identifies the point of equilibrium between rational LPs and actual LPs, corresponding to a scenario devoid of arbitrage opportunities.

Additionally, the figure integrates an earnings curve, illustrated through a magenta dashed line, to delineate the path of normalized profits. This curve commences at zero under the condition $\frac{\alpha}{\beta + p\theta} \geq \frac{p\theta-\delta}{\beta+p\theta}$, signifying the absence of speculative advantages. However, as the ratio $\frac{\alpha}{\beta + p\theta}$ diminishes below $\frac{p\theta-\delta}{\beta+p\theta}$, a monotonically decreasing trend in normalized profits is observed, relative to the initial pricing of the asset, $\alpha$.

Let us consider the limit case where $\frac{\alpha}{\beta+p\theta}$ approaches $\frac{p\theta-\delta}{\beta+p\theta}$. In this scenario, the normalized block rewards at the threshold of SBs exiting the system are represented as:

\begin{equation}
\left.\frac{\gamma}{\beta+p\theta}\right|_{ \frac{\alpha}{\beta+p\theta}\rightarrow \frac{p\theta-\delta}{\beta+p\theta}}^{\text{\tiny sb}} = 1 - \frac{\beta}{\beta+p\theta}-\frac{1}{\mathcal{N}}.
\label{eq:buyer_limit_gamma_general}
\end{equation}
Moreover, examining the limit as $\frac{\alpha}{\beta+p\theta}$ approaches an infinitesimally small value greater than zero, we find:
\begin{equation}
\left.\frac{\gamma}{\beta+p\theta}\right|_{\frac{\alpha}{\beta+p\theta}\rightarrow 0^+}^{\text{\tiny sb}} = 1 - \frac{\beta}{\beta+p\theta},
\label{eq:buyer_limit_gamma}
\end{equation}
and,
\begin{equation}
\left.\hat{\mathcal{G}}_{\text{\tiny bl}}\right|_{\frac{\theta-\delta}{\beta+p\theta}=1, \frac{\alpha}{\beta+p\theta}\rightarrow 1^+} = 0.
\label{eq:buyer_limit_G}
\end{equation}
Given that $\gamma$ must be non-negative in all cases, the condition $1-\frac{1}{\mathcal{N}} - \frac{\beta}{\beta+p\theta} \geq 0$ must hold for equation \eqref{eq:buyer_limit_gamma_general} to be valid universally. This requirement translates to $\mathcal{N} \geq 1 + \frac{\beta}{p\theta}$, signifying that for any given liquidity pool with specific setup parameters, there exists a minimum depth criterion that must be met to facilitate the swapping between non-dissipative and dissipative assets.

In Figure \ref{fig:lp_buyer}(b), we illustrate the interaction dynamics between genuine buyers and liquidity providers within the PinFi system. Drawing upon the cessation condition for genuine buyers as defined by equation \eqref{eq:gbcondition}, we identify the threshold at which genuine buyers cease their purchasing activities: when $\frac{\alpha}{\beta+p\theta}$ equals $\frac{\theta-\delta}{\beta+p\theta}$, specifically at a value of 1.3. This delineation establishes the upper limit for genuine buyer activity within the system.

The figure prominently features a shaded region, visually demarcating the parameter space within which genuine buyers are active in the market. This shaded area is crucial for understanding the conditions under which genuine buyers contribute to the system's liquidity by engaging in transactions. The demarcation serves as a guide for analyzing the impact of genuine buyers on the overall dynamics of the liquidity pool, highlighting the regions where their participation is economically viable and, thus, expected.

\subsection{Parameter Ranges and System Dynamics}

Drawing from the foundational discussions on Speculating Sellers (SSs), Speculating Buyers (SBs), Genuine Buyers (GBs), and Liquidity Providers (LPs), we can coalesce our findings into an overarching framework that elucidates the PinFi system's dynamics and equilibria. This framework is elegantly represented by the dimensionless vector:
\begin{equation}
\left(\frac{\beta}{\beta + p\theta}, \frac{p\theta-\delta}{\beta+p\theta}, \frac{\theta-\delta}{\beta+p\theta}, \frac{1}{\mathcal{N}}\right) \equiv (A, B, C, 1/\mathcal{N}).
\label{eq:normalized_vec}
\end{equation}
It is important to recognize that the parameters $A, B$, and $C$ are interconnected through the relationship $(1-p)(1-A) = p(C-B)$. 

\begin{figure}[ht]
\centering
\includegraphics[width=0.8\linewidth]{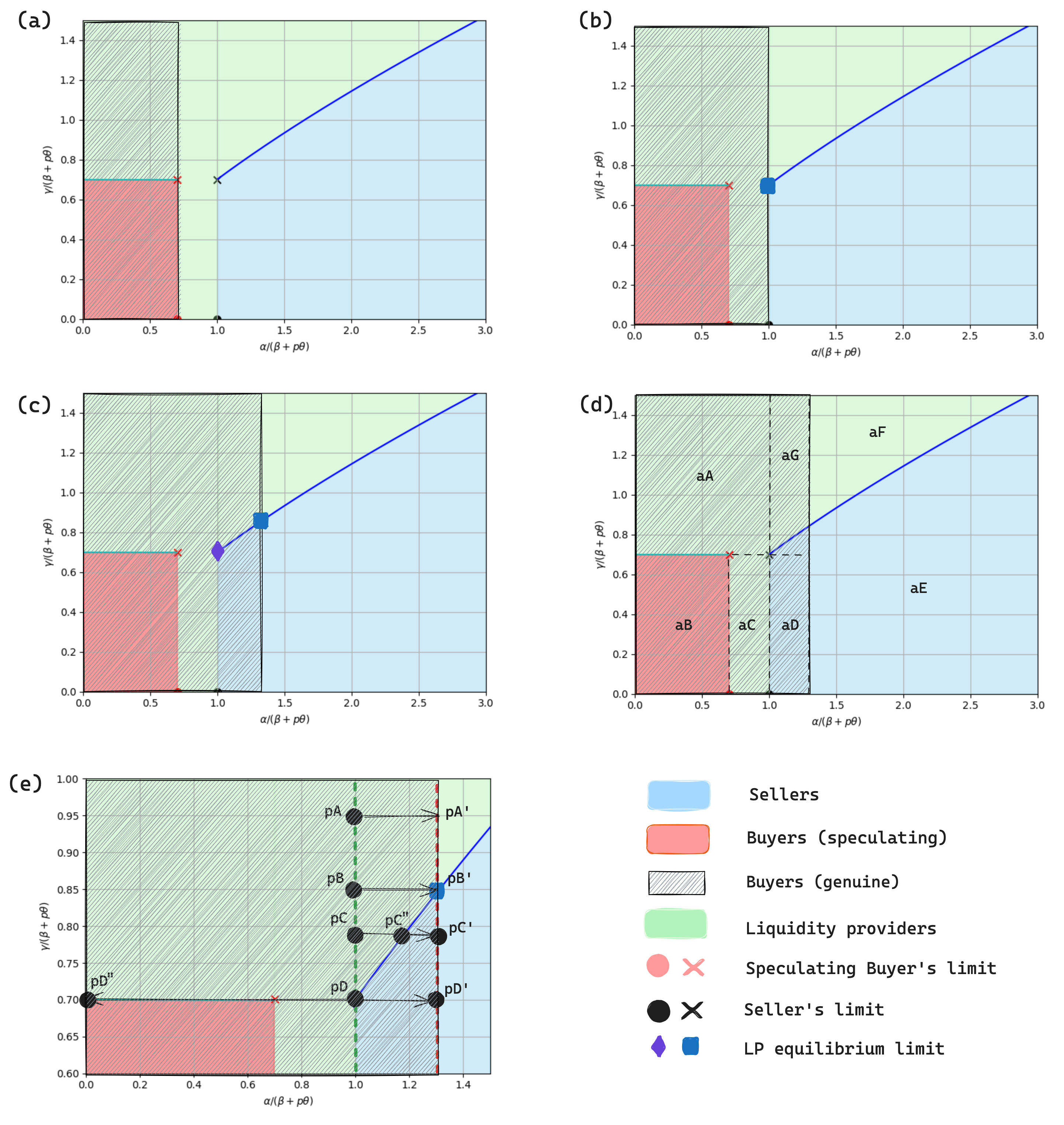}
\caption{\label{fig:dynamics}Phase Diagrams Illustrating the Dynamics and Equilibria in Dissipative Asset Pricing: Panels (a) through (c) showcase variations in the parameter $C = \frac{\theta-\delta}{\beta+p\theta}$, with values set at 0.7, 1.0, and 1.3, respectively, to depict different pricing dynamics. The diagrams incorporate constant parameters $A = \frac{\beta}{\beta+p\theta} = 0.3$ and $B = \frac{p\theta-\delta}{\beta+p\theta} = 0.7$, facilitating comparisons across scenarios. Panel (d) mirrors the configuration of panel (c) but further delineates the diagram into distinct zones, labeled aA through aG, for detailed analysis. Panel (e) provides a detailed zoom-in view of zone aG from panel (d), highlighting the intricate dynamics within this specific region.}
\end{figure}

This multifaceted vector delineates the spectrum of system behaviors, as captured in Figure \ref{fig:dynamics}. Here, phase diagrams for three discrete $C$ values—0.7, 1.0, and 1.3—are showcased, illuminating the system's behavioral variations under these conditions. Panels (a) through (c) are set against a backdrop of constant parameters: $A = \frac{\beta}{\beta + p\theta} = 0.3$, $B = \frac{p\theta-\delta}{\beta+p\theta} = 0.7$, and $\frac{1}{\mathcal{N}} = 0.001$, facilitating a clear comparison across different scenarios. Within these diagrams, the equilibria between SSs and LPs, as well as between rational LPs (considering the influence of SBs) and LPs, are demarcated by blue and red lines, respectively, providing a visual guide to understanding the equilibrium states that underpin the PinFi system's stability and functionality.

By dissecting the dynamics within the PinFi system across different parameter ranges, we obtain a nuanced view of participant roles and their viability within the system:
\begin{itemize}
\item For $C =0.7 $ (panel (a)): This scenario presents distinct divisions between sellers and buyers, including both speculating buyers (SBs) and genuine buyers (GBs), with no overlap in their operational ranges. This separation suggests that within any given initial setting of $\alpha$, sellers and buyers are mutually exclusive, leading to distinct phases of system operation:
\begin{itemize}
\item When $\frac{\alpha}{\beta + p\theta}\leq0.7$, the system can only sustain an equilibrium between liquidity providers (LPs) and buyers, precluding the presence of sellers.
\item For $\frac{\alpha}{\beta + p\theta}\in (0.7, 1.0)$,  liquidity providers are the sole viable participants, indicating a phase where the system supports neither buying nor selling activities.
\item With $\frac{\alpha}{\beta + p\theta}\geq1$, the environment becomes conducive to speculating sellers (SSs), leading to a stable configuration that, paradoxically, undermines long-term sustainability by sidelining buyers and LPs.
\end{itemize}
This analysis suggests that the parameter setting depicted in panel (a) does not foster an ideal operational environment for the PinFi system. The lack of intersecting roles within any initial $\alpha$ setting highlights a critical limitation in achieving a balanced system. The exclusion of active buying roles, especially at higher $\alpha$ values, underscores the potential for a skewed system dynamics favoring speculating sellers over other participants, thereby challenging the system's long-term sustainability and equitable participant engagement.

\item For $\frac{\theta - \delta}{\beta+p\theta} = 1.0$ (panel (b)). This parameter configuration marks a pivotal setting within the PinFi system, enabling an immediate establishment of equilibrium among the three critical roles: speculating sellers (SSs), genuine buyers (GBs), and liquidity providers (LPs). This harmonious coexistence from the system's inception indicates a well-balanced ecosystem where each participant's role is viable and contributes to the system's dynamics. Further exploration of how these dynamics unfold and impact the system's stability and sustainability is provided in subsequent analysis. This balanced interaction suggests an ideal operational state for the PinFi system, where the equilibrium fosters active participation across all roles, ensuring a dynamic yet stable marketplace for dissipative assets.

\item For $\frac{\theta - \delta}{\beta+p\theta} = 1.3$ (Panel (c)). This setting illustrates an expanded operational range for genuine buyers, extending the viable parameter space for their activity to $\frac{\alpha}{\beta+p\theta} \in [1, 1.3]$. To dissect the implications of this extended range on the system's stability and sustainability, we delve into a segmented analysis of the phase space, categorized into zones from aA to aG as depicted in Panel (d). Each zone encapsulates distinct dynamics and participant roles within the system:
\begin{itemize}
\item \textbf{Zone aA}: Dominated by liquidity providers due to the attractiveness of block rewards surpassing the potential profits from speculative selling or external market transactions. This scenario, however, is unsustainable as it sidelines other crucial market roles, particularly sellers.

\item \textbf{Zone aB}: Characterized by the mass migration of resource holders towards external selling, motivated by higher profitability compared to liquidity provision. This trend jeopardizes the liquidity pool's sustainability by diminishing the number of active LPs.

\item \textbf{Zone aC}: Despite moderate block rewards, resource holders favor the role of liquidity providers, influenced by lower external market prices. Nonetheless, the absence of sellers prevents the system from reaching a genuine equilibrium, posing sustainability issues.

\item \textbf{Zone aD}: Here, the incentive to withdraw liquidity for speculative selling overshadows the benefits of liquidity provision due to superior profitability. Although sellers and buyers can coexist, the depletion of LPs renders the AMM framework non-viable.

\item \textbf{Zone aE}: This dire situation sees a complete vacuum of buyers and liquidity providers, with all participants aiming to exploit the depleted liquidity pool for personal gain, leading to a non-sustainable state.

\item \textbf{Zone aF}: Lacks buyer presence, and rational resource holders find liquidity provision more appealing than selling, leading to a scarcity of sellers. This imbalance causes price drifts in the AMM, highlighting sustainability concerns.
\end{itemize}
Consequently, the only parameter space fostering a balanced ecosystem is encapsulated within \textbf{zone aG}. To elucidate the dynamics and assess the stability and sustainability within this specific region, Panel (e) zooms into zone aG, selecting key representative points for in-depth analysis. This approach aims to isolate conditions under which the PinFi system can achieve and maintain a harmonious equilibrium, ensuring its long-term viability and functional integrity across diverse participant roles.
\begin{itemize}
\item \textbf{pA}: At this juncture, the system lacks seller participation. Purchases by genuine buyers incrementally elevate the price of the dissipative asset until it stabilizes at point $\text{pA}'$, where genuine buyers withdraw from the market. This cessation leaves the asset price fixed at $C$, indicating a system that fails to sustain active trading or fair pricing over time, thereby questioning its sustainability.

\item \textbf{pB}: Similar to pA, this point also experiences an absence of sellers initially. However, as genuine buyers continue their purchases, the asset's price rises until reaching a critical juncture at $\text{pB}'$. $\text{pB}'$ represents a stable equilibrium where sellers, buyers, and liquidity providers each fulfill their roles, contributing to system stability. The transition to $\text{pB}'$ is contingent upon the genuine buyers' utility function, potentially leading to prolonged periods of price adjustment between $\text{pB}$ and $\text{pB}'$.

\item \textbf{pC}: Should genuine buyers incrementally increase the asset price to $\text{pC}''$, the system stabilizes at this higher valuation. If the asset pricing in the liquidity pool surpasses $\text{pC}''$, some LPs may opt to assume the role of sellers, capitalizing on arbitrage opportunities. This shift aims to rebalance the asset's price back to $\text{pC}''$, maintaining equilibrium.

\item \textbf{pD}: This point is inherently stable, supporting the coexistence of sellers, buyers, and LPs. An increase in the liquidity pool's price beyond pD prompts some LPs to transition into sellers, seeking higher profits through arbitrage. This mechanism ensures the system's return to the equilibrium state at pD, underlining its resilience.
\end{itemize}
The identification of a potentially stable range along the blue line from pD to $\text{pB}'$ offers insight into the system's capacity to reach equilibrium. However, the introduction of a hypothetical scenario involving an irrational resource-holder, intent on unloading all their resources into the liquidity pool for rapid gain, introduces a significant stress test for the system's resilience. This action poses the risk of drastically driving down the price, potentially destabilizing the equilibrium achieved within the stable range. To assess the system's robustness against such extreme behavior, we turn to the implications of equations \eqref{eq:seller_limit_gamma} and \eqref{eq:buyer_limit_gamma}:
\begin{equation}
\left.\frac{\gamma}{\beta+p\theta}\right|_{\frac{\alpha}{\beta+p\theta}\rightarrow 0^+}^{\text{\tiny sb}} < \left.\frac{\gamma}{\beta+p\theta}\right|_{\frac{\alpha}{\beta+p\theta}\rightarrow 1^+}^{\text{\tiny ss}}.
\end{equation}
This inequality suggests that, by strategically setting the initial $\gamma$ within the range between pD and $\text{pB}'$, the system can shield itself against the destabilizing effects of irrational selling actions. In other words, even if an irrational seller attempts to flood the market with resources, aiming to drive the price to zero, the structure and settings of the system ensure that liquidity providers are incentivized to maintain their stakes in the liquidity pool rather than diverting resources to external markets. This built-in safeguard reinforces the system's stability by ensuring liquidity providers' continued participation, even under potentially disruptive market actions. Thus, this analysis not only highlights the resilience of the PinFi system in maintaining equilibrium across a variety of scenarios but also underscores its capacity to withstand extreme situations, ensuring sustainability and the integrity of pricing mechanisms within the decentralized finance landscape.

So, the normalized block rewards should be bounded within a specific range. This range is mathematically expressed as:
\begin{equation}
\frac{\gamma}{\beta+p\theta} \in \left[\frac{\gamma_{\text{\tiny lower}}}{\beta+p\theta}, \frac{\gamma_{\text{\tiny upper}}}{\beta+p\theta}\right],
\end{equation}
where the lower and upper limits of normalized block rewards are defined by:
\begin{equation}
\frac{\gamma_{\text{\tiny lower}}}{\beta+p\theta} = 1+\frac{1}{\mathcal{N}}-A, \,\,\,\, \frac{\gamma_{\text{\tiny upper}}}{\beta+p\theta} = \left(1+\frac{1}{\mathcal{N}}\right)\frac{C-1}{\ln C}-A.
\end{equation}
These bounds ensure that block rewards are calibrated to foster an environment where liquidity providers are motivated to remain within the system, rather than seeking potentially higher but more volatile returns through speculative selling or external market transactions. This calibration hinges on setting an $\alpha$ that falls between two critical values: $\alpha_{\text{\tiny lower}} = \beta + p\theta$ and $\alpha_{\text{\tiny upper}} = \theta - \delta$. This parameterization ensures the system's equilibrium and sustainability when the condition $\alpha_{\text{\tiny upper}} > \alpha_{\text{\tiny lower}}$ holds true, which simplifies to the condition $\beta +\delta < (1-p)\theta$.

Exploring the upper bound of $\gamma$ offers valuable insights, particularly in scenarios where the transaction fee for users, denoted as $\delta$, is zero. Under this condition, equation \eqref{eq:normalized_vec} simplifies the expression for $C$ to $C = (1-A)/p$. Therefore, with known values of $A$ and $p$, the lower and upper bounds of $\gamma$ can be precisely calculated.

\end{itemize}

\section{Summary and Ongoing Work}
\label{ongoingwork}

This paper delves into the PinFi protocol and its extensive implications, illustrating that under common conditions and assuming participant integrity, it can achieve three distinct equilibria: selling-liquidity, buying-liquidity, and selling-buying-liquidity. Our analysis delineates vital parameter ranges essential for the system's stability and sustainability, thereby reducing arbitrage opportunities and ensuring fair pricing for dissipative assets.

However, due to the economic system's inherent complexity, this study is confined to a narrower scope, focusing primarily on the general dynamics within the system. Specifically, we examine the interactions between genuine buyers and liquidity providers (GB-LP), speculating buyers and liquidity providers (SB-LP), and speculating sellers and liquidity providers (SS-LP), as indicated by the black lines in Figure \ref{fig:scheme}(c). Meanwhile, other potential interactions, marked by green dotted lines, are not explored in depth. These additional interactions could offer further insights into the system's susceptibility to chaos or bifurcations, thereby informing strategies for more effective system management. However, incorporating these aspects would necessitate additional assumptions about participant behavior beyond the rational actor model employed in our current analysis. To maintain focus, we reserve an in-depth examination of these dynamics for future work.

Moreover, to validate our theoretical findings, detailed Monte Carlo simulations are necessary. Such simulations could relax the assumptions regarding participant behavior, making our model more reflective of real-world conditions and enabling a more precise identification of critical parameter ranges.

Lastly, given our focus on the decentralized pricing of dissipative assets, further analysis incorporating on-chain simulations is essential. This would account for the blockchain's discrete nature, adding another layer of realism to our study and enriching our understanding of decentralized financial mechanisms.

\section*{Acknowledgements}
This document stands as a testament to the dedication of our team members, whose tireless collaboration was essential to its realization. 

\bibliographystyle{alpha}
\bibliography{sample}

\end{document}